\documentclass[useAMS,usenatbib]{mn2e}
\usepackage{graphicx}  
\usepackage{psfig}  
\newcommand{\beq}{\begin{equation}}  
\newcommand{\eeq}{\end{equation}}  
\newcommand{\beqa}{\begin{eqnarray}}  
\newcommand{\eeqa}{\end{eqnarray}}  
\newcommand{\sds}{\dot{\Sigma}_{\rm SFR}}  
\newcommand{\apj}{ApJ}
\newcommand{\apjl}{ApJL}
\newcommand{\apjs}{ApJS}
\newcommand{\aap}{A\&A}
\newcommand{\nat}{Nature}
\newcommand{\mnras}{MNRAS}
\newcommand{\physrep}{Phys.~Rep.}

\newcommand{\araa}{ARAA}
\def\lesssim{\mathrel{\raise1.5pt\hbox{\rlap{\hbox{\lower5pt\hbox{$\sim$}}}\hbox{$<$}}}}
\def\gtrsim{\mathrel{\raise1.5pt\hbox{\rlap{\hbox{\lower5pt\hbox{$\sim$}}}\hbox{$>$}}}}

\title[Radio Emission from Supernova Remnants] {Radio Emission from
  Supernova Remnants: Implications for Post-Shock Magnetic Field Amplification
  \& the Magnetic Fields of Galaxies}

\author[Thompson, Quataert, \& Murray]
{Todd A.~Thompson$^1$, Eliot Quataert$^2$, \& Norman Murray$^{3,4}$\\ 
\noindent$^1$Department of Astronomy and Center for Cosmology \& Astro-Particle Physics,
The Ohio State University, Columbus, Ohio 43210 \\ 
$^2$Astronomy Department \& Theoretical Astrophysics Center, 
601 Campbell Hall, The University of California, Berkeley, CA 94720\\  
$^3$Canadian Research Chair in Astrophysics\\
$^4$Canadian Institute for Theoretical Astrophysics, 
The University of Toronto, 60 St.~George Street, Toronto 
Ontario M5S 3H8}

\begin{document}

\date{Accepted . Received ; in original form }

\pagerange{\pageref{firstpage}--\pageref{lastpage}} \pubyear{}

\maketitle

\label{firstpage}

\setlength{\baselineskip}{10.75pt}    

\begin{abstract}  

\setlength{\baselineskip}{10.75pt}
    
Using observational data from the literature, we show that the non-thermal 
radio luminosity ($L$) of supernova remnants (SNRs) is a strong function of 
the average gas surface density ($\Sigma_g$) of the galaxy in which the 
remnants reside, from normal spirals to dense luminous starbursts. Our 
result supports the interpretation of the radio sources in M82 and Arp 220 
as normal SNRs, and not ``radio'' supernovae.  We combine a simple theory 
for electron cooling in SNRs with their observed radio luminosities to 
estimate the remnant magnetic field
strength ($B_{\rm SNR}$): the correlation between $L$ and $\Sigma_g$
implies that $B_{\rm SNR}$ also increases with $\Sigma_g$.  We explore
two interpretations of this correlation: (1) $B_{\rm SNR}$ is
generated by post-shock magnetic field amplification, with $B_{\rm
  SNR}^2 \propto \Sigma_g$ and (2) $B_{\rm SNR}$ results from
shock-compression of the ambient interstellar medium (ISM) magnetic
field ($B_{\rm ISM}$), with $B_{\rm ISM}$ being larger in denser
galaxies.  We find that shock compression is, on average, sufficient
to produce the observed radio emission from SNRs in the densest
starburst galaxies; amplification of post-shock magnetic fields is not
required.  By contrast, in normal spirals modest post-shock field
amplification in some remnants (a factor of $\sim{\rm few}-10$) is
consistent with the data; we find tentative evidence that both the
Alfv\'en speed within SNRs and the ratio of $B_{\rm SNR}^2/8\pi$ to
the post-shock pressure (``$\epsilon_B$'') are constant in
SNRs from galaxy to galaxy.  We discuss observational tests that can be used to
more definitively distinguish between these two interpretations of the
radio luminosities of SNRs.  Regardless of which is correct, the radio
emission from SNRs provides an upper limit to $B_{\rm ISM}$ that is
independent of the minimum energy assumption.  For the densest
starbursts, the magnetic energy density in the ISM is below the
total ISM pressure required for hydrostatic equilibrium; thus magnetic
fields are not dynamically important on the largest scales in
starbursts, in contrast with spiral galaxies like our own.  This
dichotomy may have implications for galactic dynamo theory.

\end{abstract}


\section{Introduction}  
\label{section:intro} 

The magnetic energy density of the Galaxy is in rough equipartition
with both the cosmic ray and the turbulent energy densities.  These
components of the ISM combine to produce a midplane pressure
sufficient to balance the self-gravity of the Galactic disk and thus
to establish hydrostatic equilibrium (Boulares \& Cox 1990).
Moreover, direct probes of magnetic fields in dense star-forming
regions within the Galaxy indicate that they may be dynamically
important (Crutcher 1999).  Because of the potential importance of
magnetic fields for star formation on small scales, at and below the
scale of giant molecular clouds, and for the structure and
self-regulation of galaxies on their largest scales, it is of
considerable interest to understand how the average magnetic field
strength varies from normal galaxies to starbursts, where the ISM
conditions are considerably different.

A rough upper limit to the average magnetic field strength in a galaxy
can be constructed by assuming that the field is dynamically
comparable to gravity.   For a thin self-gravitating
gas-dominated disk of average gas surface density $\Sigma_g$, this
``equipartition'' field strength is \beq B_{\rm eq}\approx 2 \pi
\,\sqrt{2G} \,\Sigma_g\approx\,2.3 \,\Sigma_g\,\,\,{\rm mG},
\label{beq}
\eeq where $\Sigma_g$ is measured in g cm$^{-2}$.  $B_{\rm eq}$
provides an upper limit to $B$ on galactic scales because for
$B>B_{\rm eq}$ the field is buoyant and escapes the host galaxy
(Parker 1966).  For galaxies like the Milky Way, in which the total
surface density, $\Sigma_{\rm tot}$, is dominated by stars, $B_{\rm
  eq}$ in equation (\ref{beq}) should be multiplied by a factor of
$(\Sigma_{\rm tot}/\Sigma_g)^{1/2} \, (\approx 3)$.
The measured gas surface densities of galaxies vary by more than 4 decades
(e.g., Kennicutt 1998, hereafter K98). Over this range, equation (\ref{beq}) 
implies a maximum field strength of $B_{\rm eq}\approx15$\,$\mu$G for the
Milky Way ($\Sigma_g\approx2\times10^{-3}$\,g cm$^{-2}$; $\Sigma_{\rm
tot}/\Sigma_g\approx10$) and $\sim20$\,mG in galaxies like Arp 220
that anchor the high-density end of the Schmidt Law with
$\Sigma_g\approx10$\,g cm$^{-2}$ and $\Sigma_{\rm
tot}/\Sigma_g\approx1$ (Downes \& Solomon 1998).

Estimates of magnetic field strengths in galaxies are traditionally
limited to the ``minimum energy'' assumption of Burbidge (1956), which
posits an equality between the magnetic and cosmic ray energy
densities (see Longair 1994; Beck \& Krause 2005).  This approximation
works well in the Galaxy and other normal spirals where the field
strength is also nearly equipartition in the sense of equation
(\ref{beq}), but the minimum energy assumption likely underestimates
the field strengths in starburst galaxies because of strong cosmic ray
electron cooling (Condon et al.~1991; Chi \& Wolfendale 1993; Thompson
et al.~2006; hereafter [T06]).  

One way to see that the minimum energy estimate must fail in
ultra-luminous infrared galaxies (ULIRGs) like Arp 220 is to note that
the inverse Compton (IC) cooling time for radio-emitting cosmic ray
electrons is $t_{\rm IC}\sim10^4$\,yr (e.g., Condon et al.~1991).  The
minimum energy estimate for the magnetic field strength yields a
synchrotron cooling timescale at GHz frequencies that is $\sim 10$
times longer, and yet Arp 220 lies on the FIR-radio correlation
together with essentially all star-forming galaxies (Condon 1992; Yun
et al.~2001), which have synchrotron cooling times somewhat
shorter than IC cooling times (e.g., Fig.~2 of T06).  These facts 
imply that the true magnetic field strength in
ULIRGs is considerably larger than the minimum energy estimate, and
that rapid electron cooling invalidates the assumptions upon which
that estimate is predicated.

This argument can be generalized and used to construct an empirically
derived {\it minimum} magnetic field strength in galaxies by equating
the magnetic energy density $U_B$ and the energy density in starlight
$U_{\rm ph}$.  Because the ratio $U_B/U_{\rm ph}$ determines the ratio
of synchrotron to IC cooling, and because normal galaxies and ULIRGs
lie on {\it both} the FIR-radio correlation and the Schmidt Law, $U_B
\equiv B^2/8\pi$ must be larger than (or a constant fraction of)
$U_{\rm ph}=F/c=\epsilon\sds c$, where $\epsilon$ is a stellar
IMF-dependent constant, $F$ is the flux, and $\sds$ is the star
formation rate per unit area.\footnote{In principle, a factor of the
dust optical depth should be applied to $U_{\rm ph}$ in the densest
starburst galaxies, if the cosmic rays are co-spatial with the dense
molecular gas that is optically thick even in the FIR (see, e.g.,
Thompson et al.~2005).  However, this correction is uncertain and for
the purposes of constructing a lower limit to $B$ based on $U_{\rm
ph}$, this expression suffices.}  Thus, the average magnetic field
strength must exceed \\
\\
\beqa B_{\rm ph}=(8\pi\epsilon\sds
c)^{1/2}&\approx&0.3\,\Sigma_g^{0.7}\,\,{\rm mG} \nonumber \\
&\approx&1.0\,\Sigma_g^{0.85}\,\,{\rm mG},
\label{bph}
\eeqa where the numerical approximations follow from the observed Schmidt
Laws of K98 (top) and Bouch\'e et al.~(2007) (bottom).  
$B_{\rm ph}$ is a lower limit to the mean field strength in galaxies
because if $U_B \lesssim U_{\rm ph}$, variations in $U_{B}/U_{\rm ph}$
from galaxy to galaxy would likely introduce scatter and non-linearity
into the FIR-radio correlation (T06).  For the Galaxy, $B\gtrsim
B_{\rm ph}\approx4$\,$\mu$G and for Arp 220, $B\gtrsim B_{\rm
ph}\approx1$\,mG.

In this paper, we argue that radio observations of supernova remnants
(SNRs) provide a complimentary probe of the ISM magnetic field
strength in star-forming galaxies.  In \S\ref{section:radio} we
describe a simple synchrotron cooling model for the radio luminosity
of an individual SNR.  The radio flux depends critically on the
magnetic field in the SNR, which is somewhat uncertain.  At a minimum,
a SNR must contain shock-compressed ISM magnetic field, which is
constrained (on average) by the upper and lower limits of equations
(\ref{beq}) and (\ref{bph}), respectively.  However, the magnetic
field in SNRs could be much stronger than the shock compressed ISM
field if there is significant amplification in the post-shock plasma.
We compare the remnant magnetic field strengths derived from the model
presented in \S\ref{section:radio} with our expectations from
equations (\ref{beq}) and (\ref{bph}), to set limits on the importance
of field amplification in SNRs. We further argue that the radio
emission from SNRs provides an {\it upper limit} on the ambient ISM
field in star-forming galaxies.  In \S\ref{section:results} we show
that the luminosity of SNRs is strongly correlated with the average
gas surface density of the galaxy in which they reside.  Using the
model of \S\ref{section:radio}, we invert these observations to
constrain the remnant and ambient ISM magnetic field in this sample of
galaxies.  We discuss our results in \S\ref{section:discussion},
focusing on the relative importance of shock compression of ISM field
versus post-shock field amplification.

\section{Magnetic Fields in Supernova Remnants}
\label{section:radio}

To estimate the radio luminosity of a SNR, we assume that a fraction
$10^{-2} \xi$ of the supernova kinetic energy ($E_{\rm 51}=E_{\rm
SN}/10^{51}$ ergs) is supplied to primary cosmic ray electrons in the
shock and that the accelerated electrons radiate synchrotron in a
magnetic field of strength $B_{\rm mG}=B/{\rm mG}$.  We further assume
that the electron particle spectrum is flat, with a power-law index of
$p = 2$ ($n(\gamma) \propto \gamma^{-p}$), as is expected
theoretically for strong shocks (e.g., Blandford \& Eichler 1987) and
is observed {\it in situ} in some SN remnants (e.g., Aharonian et
al.~2005; Brogan et al.~2005).  A value of $\xi \approx 1$ is required
to explain the integrated radio flux from star-forming galaxies, i.e.,
the FIR-radio correlation (V\"olk 1989; T06).  Given the small scatter in the
FIR-radio correlation (Yun et al.~2001), $\xi \approx 1$ is uncertain
at the factor of $\lesssim 2$ level when averaged over many SNe in a
galaxy.  With these assumptions, the radio luminosity is simply (van
der Laan 1962) \beq \nu L_\nu \approx {\xi E_{\rm SN} \over 2
\ln[\gamma_{\rm max}]\, t_{\rm syn}} \approx 3 \times 10^{35} \xi \,
E_{51} \,\nu_{\rm GHz}^{1/2} \,B_{\rm mG}^{3/2} \, \,\,{\rm ergs \,
\,\,s^{-1}},
\label{lrad} 
\eeq where $t_{\rm syn} \approx 3 \times 10^{4}\,\,B_{\rm
mG}^{-3/2}\,\nu_{\rm GHz}^{-1/2}\,\,\,{\rm yr}$ is the synchrotron
cooling timescale for electrons radiating at $\nu_{\rm GHz}=\nu/{\rm
GHz}$ and where we have assumed that $t_{\rm syn} \gtrsim t_{\rm
exp}$, where $t_{\rm exp}$ is the expansion time of the remnant.  In
the last estimate in equation (\ref{lrad}) we have assumed
$\gamma_{\rm max}=10^6$; variations in $\gamma_{\rm max}$ from
$10^{4}-10^{8}$ affect this estimate by only a factor of $\sim1.5$.

Equation (\ref{lrad}) is an estimate of the radio emission produced by
the interaction between a SN and the ambient ISM, i.e., the emission
from a SNR (e.g., Shlovskii 1960).  The radio emission is expected to
peak at the Sedov time and persist until the remnant cools radiatively
(see the discussion of these timescales in
\S\ref{section:discussion}).  This time evolution is in contrast to
{\it radio supernovae} which peak and decay after a month to few-years
and which likely result from the interaction between a SN shock and
the immediate circumstellar medium and pre-SN ejecta (e.g., Chevalier
1982; Weiler et al.~2002).  Nearly all of the radio remnants
discovered in M82 and Arp 220 that we discuss below do not vary
appreciably on decade timescales (Kronberg et al.~2000; Rovilos et
al.~2005).  Rather than interpret these bright sources as {\it bona
fide} radio SNe, we thus interpret them as normal SNRs.

Given an observed radio flux density, equation (\ref{lrad}) can be
inverted to estimate the magnetic field strength in the SNR, $B_{\rm
SNR}$.  Assuming that $B_{\rm SNR}$ is ambient ISM field compressed by
the SN shock by a factor of $f\approx3$ (appropriate for
randomly-oriented fields; V\"olk et al.~2002 advocate $f\approx6$ for
young remnants), we obtain a simple estimate of the ISM magnetic field
strength in terms of the SNR flux density: \beq
\hspace*{-.1cm}
B_{\rm ISM}\approx 3\,
\left(\frac{3}{f}\right)
\hspace*{-.1cm}\left(\frac{D}{10\,{\rm Mpc}}\right)^{\hspace*{-.05cm}4/3}
\hspace*{-.1cm}\left(\frac{S_\nu}{{\rm mJy}}\right)^{\hspace*{-.05cm}2/3}
\hspace*{-.23cm}\xi^{-2/3} \,
\hspace*{-.1cm} \nu_{\rm GHz}^{1/3} \, \, {\rm mG},
\label{bsnr}
\eeq 
equivalent to $B_{\rm ISM}f\approx B_{\rm SNR}$.
Equation (\ref{bsnr}) is most readily interpreted as an {\it upper
limit} on the ISM magnetic field given a set of SNRs with flux density
$S_\nu$ because (1) $f$ may be as large as $\sim7$ and (2) turbulence
and instabilities can in principle amplify the post-shock magnetic
field.  The SNR field may thus be larger than $\sim f B_{\rm ISM}$,
but it is unlikely to be smaller.\footnote{If SNe systematically
sample only low magnetic field regions of a galaxy, $B_{\rm ISM}$ inferred
from SNRs may not be the same as the volume averaged magnetic field
strength.  This possibility cannot be ruled out, but we regard it as
unlikely.} 

\begin{figure}
\centerline{\psfig{file=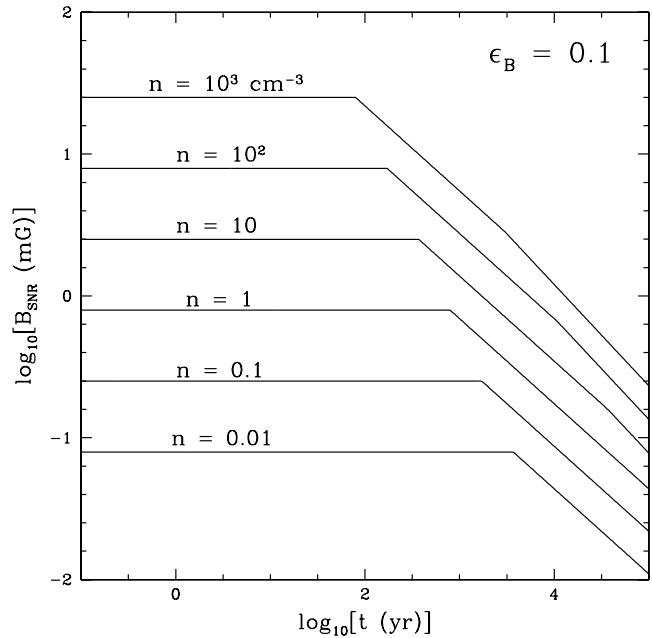,width=9cm}}
\caption{The time evolution of the post-shock
magnetic field strength, assuming $B^2/8\pi=\epsilon_B P_{\rm sh}$, 
where $P_{\rm sh}$ is the post-shock thermal pressure.   A  
$10^{51}$ ergs supernova shock and $\epsilon_B = 0.1$ are assumed in 
all cases. With these assumptions, the post-shock magnetic field is 
strongest during the free expansion phase, and declines during the 
Sedov-Taylor and snow-plow phases.
\label{fig:bshock}}
\end{figure} 

\begin{figure*}
\centerline{\psfig{file=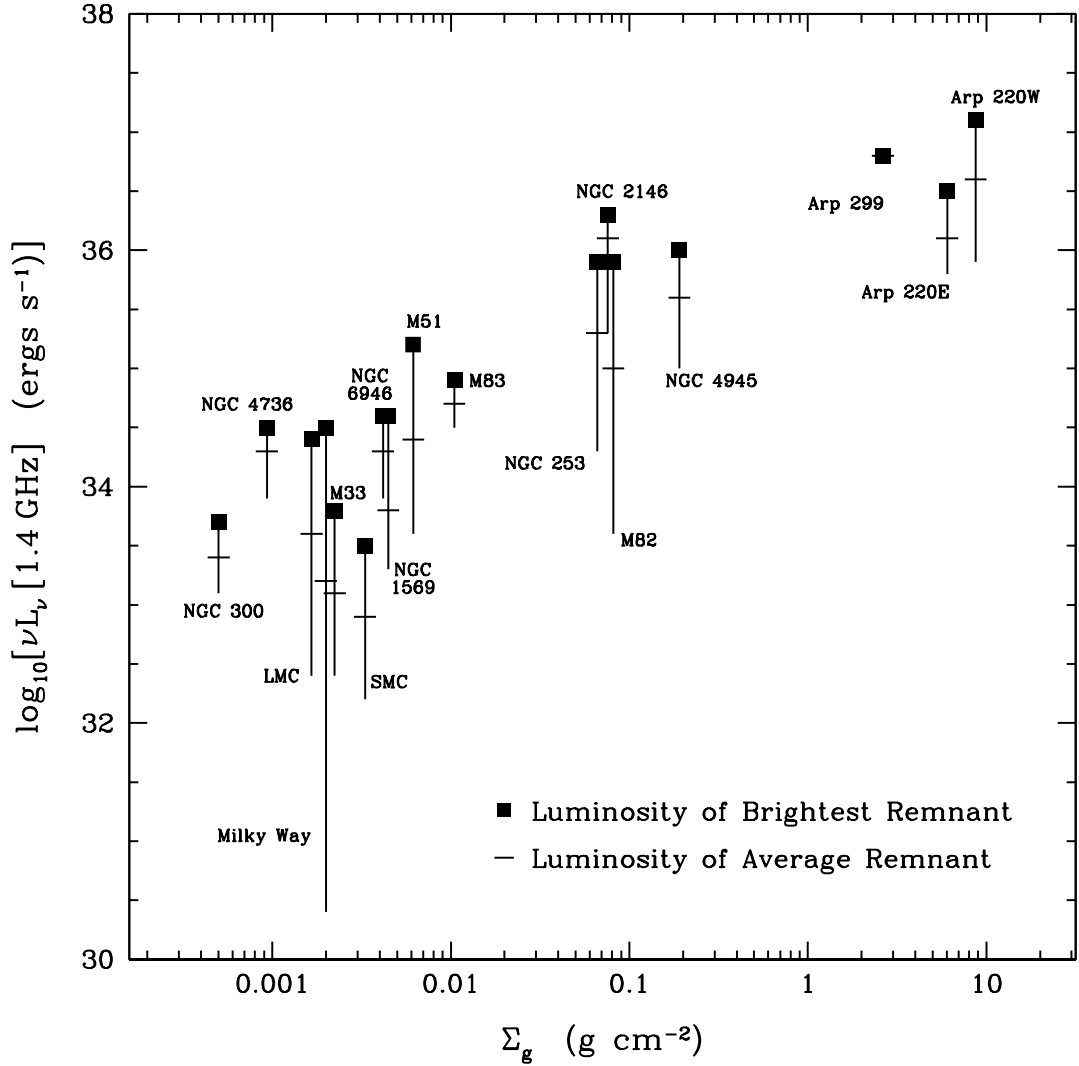,width=15cm}}
\caption{SNR radio luminosity versus average gas
surface density, $\Sigma_g$, for the sample of star-forming galaxies
in Table \ref{table:snr}.  The filled square and horizontal bar show the
maximum and average SNR luminosity for each system.  The vertical bar
extends to the lowest SNR luminosity observed.  A strong correlation is 
present in the data.  A simple linear least squares 
fit to the average SNR luminosities
gives $\nu L_\nu\propto\Sigma_g^{0.85}$;  fitting to only the brightest 
remnants gives $\nu L_\nu\propto\Sigma_g^{0.78}$ 
(see \S\ref{section:results}; eq.~\ref{fit}).  Note that although
for some galaxies many SNRs are observed (e.g., 
M33 with 51 SNRs), others, like 2146, have only a handful,
or in the case of Arp 299, a single confirmed SNR. 
For a similar figure, see Hunt \& Reynolds (2006).
\label{fig:meanl}}
\end{figure*} 

The assumptions made in deriving equation (\ref{bsnr}) include that
$t_{\rm syn} \gtrsim t_{\rm exp}$ (weak cooling) and that $p = 2$.
Together, these assumptions imply that the synchrotron spectral index
should be $\alpha=(p-1)/2=1/2$ ($S_\nu\propto\nu^{-\alpha}$).  For
$p\ne2$, and fixed total energy injected in cosmic ray electrons, the
normalization of equation (\ref{bsnr}) increases, but modestly: for
$p=2.6$, $B_{\rm ISM}$ is larger by a factor of $\approx1.5$ and
$B_{\rm ISM}\propto S_\nu^{5/9}$.  The spectral indices of GHz radio
emission from SNRs range from $\alpha\approx0.5-1$ (see the refs.~in
Table \ref{table:snr}).  Diversity in $\alpha$ may signal diversity in
$p$, evolutionary effects, that the assumption of weak cooling is
violated, or some combination.  Additionally,  note also that  
$\xi\approx1$ is not valid during the early
free-expansion phase of SNR evolution.

A further assumption made in deriving equation (\ref{bsnr}) is that
the supernova remnants are not radiative.  If they were, $f$ could be
arbitrarily high.  Indeed, Chevalier \& Fransson (2001) argued that
the radio point sources observed in M82 and Arp 220 are radiative, but
otherwise normal, supernova remnants (i.e., not ``radio'' SNe).  As we
show below and as we discuss in more detail in
\S\ref{section:discussion}, the radio luminosities of the remnants in
starburst galaxies are consistent with compression of the ambient ISM
magnetic field by an adiabatic shock, perhaps together with modest
post-shock field amplification.  Much larger $f$, as would be implied
if the remnants were radiative, is inconsistent with independent
constraints on the ambient ISM magnetic fields in these systems.

To provide some quantitative context for comparing the strength of the
magnetic field in SNRs produced by flux freezing with that produced by
possible amplification in the post-shock plasma, Figure
\ref{fig:bshock} shows the post-shock magnetic field strength in SNRs
as a function of time assuming that the post-shock remnant magnetic
energy density is a factor $\epsilon_B$ times the post-shock pressure
(as is often assumed in models of radio supernovae and gamma-ray burst
afterglows; e.g., Reynolds \& Chevalier 1981); we consider a range of
ambient densities $n$, from low densities appropriate to SNRs in the
Milky Way to high densities appropriate to SNRs in Arp 220.  For these
calculations we have used the analytic approximations for the
evolution of SNRs developed by Draine \& Woods (1991), which include
both the Sedov-Taylor and pressure-driven snow-plow phases.  The
post-shock magnetic field is the strongest during the free expansion
phase and decreases during the Sedov-Taylor phase and later as the
post-shock plasma pressure decreases.  We note, however, that the
appropriate value for $\epsilon_B$ is quite uncertain, and thus it is
difficult to estimate from first principles the importance of
post-shock field amplification (see Riquelme \& Spitkovsky 2009 for a
recent study).

\begin{figure*}
\centerline{\psfig{file=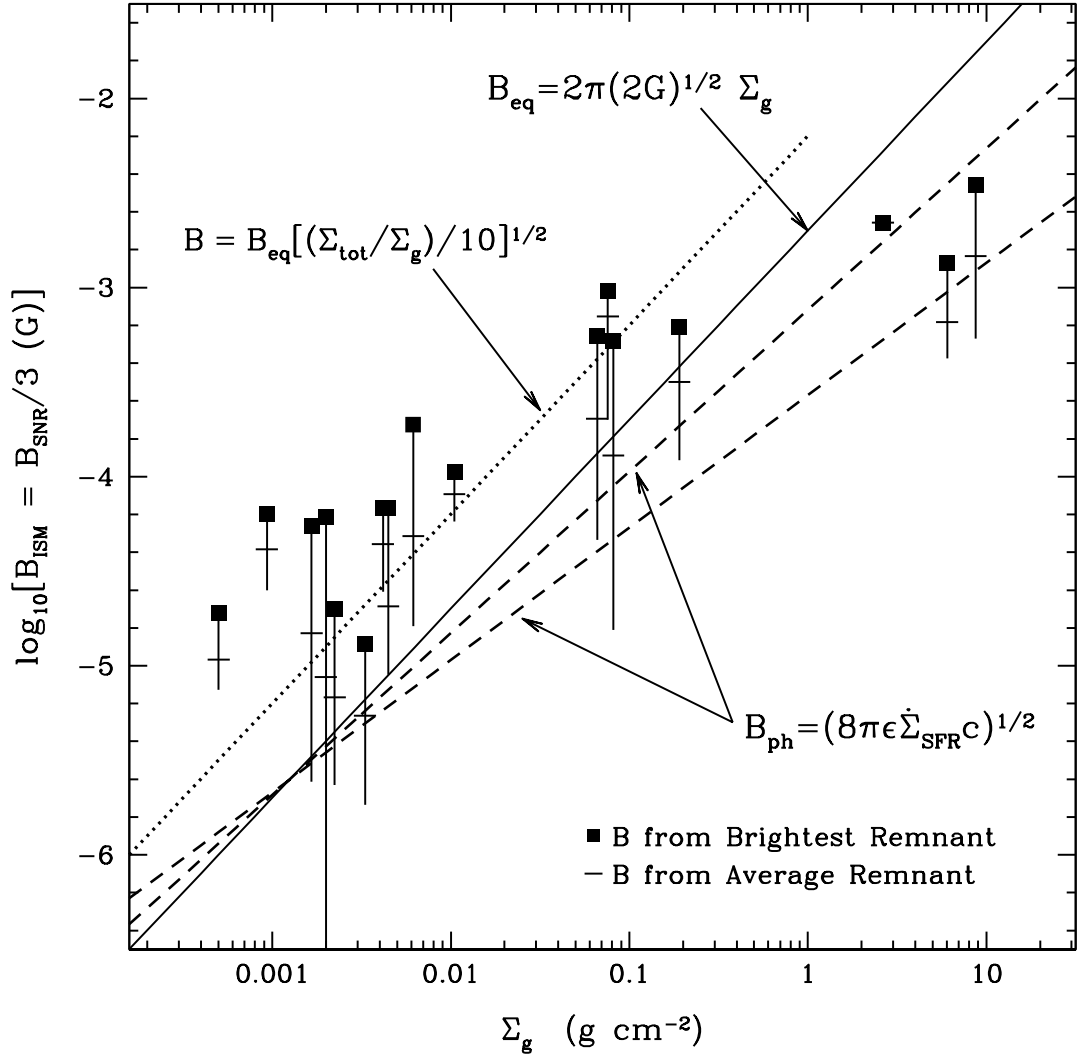,width=15cm}}
\caption{Inferred ISM magnetic field strength as function of
  $\Sigma_g$ for the galaxies in Table \ref{table:snr} (from
  eq.~\ref{bsnr} with $f = 3$), assuming that shock compression of
  ambient ISM field determines the magnetic field strength in SNRs
  ($B_{\rm SNR}$); alternatively, this plot can be interpreted as
  showing $B_{\rm SNR}/f$.  As in Figure \ref{fig:meanl}, the filled
  square and horizontal bar show the maximum and average value of $B$
  from the brightest and average SNR, for each system.  The vertical
  bar extends to the lowest SNR luminosity observed.  The thick solid
  line is the maximum magnetic field estimate for a gas-dominated
  self-gravitating disk, $B_{\rm eq}$ (eq.~[\ref{beq}]).  The dotted
  line includes the correction discussed after eq.~(\ref{beq}) for
  disks that are not gas-dominated ($\Sigma_{\rm tot}/\Sigma_g=10$).
  The dashed lines show the minimum required magnetic field strength
  for consistency with the FIR-radio correlation (see eq.~\ref{bph}),
  assuming the average Schmidt Law from K98 (shallower index) and
  Bouch\'e et al.~(2007) (steeper index).  An unweighted linear
  least-squares fit to $B_{\rm ISM}$ or $B_{\rm SNR}$, inferred from
  the average luminosity SNR (horizontal lines on each bar), gives
  $B\propto\Sigma_g^{0.56}$, whereas a fit to the brightest SNRs
  (highest inferred field strengths, filled squares) gives
  $B\propto\Sigma_g^{0.52}$ (compare with Fig.~\ref{fig:meann}). 
  \label{fig:mean}}
\end{figure*} 

\section{Results}
\label{section:results}

A large sample of radio SNRs is available in the literature.  These
SNRs reside in a variety of galaxies that span the full range of both
the FIR-radio correlation and the Schmidt Law.  In general, the
remnants are identified using the VLA, MERLIN, or VLBI, and are distinguished
from compact HII regions by their steep non-thermal spectra
($\alpha\gtrsim0.5$). The few ``radio SNe'' proper that appear in the 
compilations of SNe presented in the references in Table \ref{table:snr}
are excluded.  These include  remnant 41.95+57.5 in
M82 from Kronberg et al.~(1985) (see Kronberg et al.~2000)
and SN 2000ft in NGC 7469 (Colina et al.~2001; Alberdi et al.~2006).
Background objects such as AGN behind nearby galaxies like NGC 5194
are a contaminant to the SNR sample (see Table \ref{table:snr}).

Figure \ref{fig:meanl} shows the luminosity of observed SNRs as a
function of the average gas surface density $\Sigma_g$ of the galaxy
in which they reside.  For simplicity, all fluxes have been scaled to
a frequency of 1.4\,GHz assuming $\alpha=0.5$, even in those cases
where $\alpha$ is known empirically in some frequency range.
References, assumed distance, number of remnants, maximum and average 
SNR luminosity, and the sensitivity, frequency, and resolution of the observations 
are given in Table \ref{table:snr}.  
For all of the galaxies we consider, the remnants account for just a few to ten
percent of the total non-thermal continuum.  Each vertical line in
Figure \ref{fig:meanl} is a separate galaxy (or, for Arp 220,
component).  The filled square shows $\nu L_\nu$ for the brightest
single SNR.  The line for each system extends to the lowest luminosity
remnant identified.  The horizontal line shows the average luminosity
of the observed SNR distribution.  A similar figure was constructed by
Hunt \& Reynolds (2006).

A strong correlation is evident between $\nu L_\nu$ and $\Sigma_g$.  
A simple power-law fit to the average SNR luminosities 
(horizontal bars) gives 
\beq
\log_{10}\left[\frac{\nu L_\nu({\rm 1.4\,GHz})}{{\rm ergs\,\,s^{-1}}}\right]\approx36.03
+0.85\log_{10}\left[\frac{\Sigma_g}{{\rm g\,\,cm^2}}\right].
\label{fit}
\eeq A similar fit is found for the brightest SNRs (filled squares),
but with a normalization of 36.48 and a slope of 0.78.  It is also
clear that the {\it lowest} SNR luminosities exhibit a correlation
with $\Sigma_g$ that is qualitatively similar to that of the average
and brightest SNRs.

The latter raises the concern that the results of Figure
\ref{fig:meanl} may represent a selection bias rather than a physical
effect. The correlation between the {\it maximum} SNR luminosity and
$\Sigma_g$ in Figure \ref{fig:meanl} is the most secure and is not
subject to an obvious bias.  In particular, although faint remnants in
high-$\Sigma_g$ galaxies (which are rarer and thus tend to be more
distant) would be undetectable, bright remnants could be readily seen
in nearby low-$\Sigma_g$ systems, but are not.  On the other hand, the
correlations between the average and lowest remnant luminosity and
$\Sigma_g$ are more subject to the obvious bias that faint remnants
are difficult to detect in all galaxies.  One might anticipate that
this bias would be particularly severe for higher $\Sigma_g$ galaxies
and could account for the absence of low luminosity remnants in these
systems.  One indication that this effect does not significantly affect 
our calculation of the average SNR luminosity is that nearly all 
of the detected remnants in M82, at a distance of
3.6\,Mpc, are more luminous than even the brightest remnant in NGC 300
at a distance of 1.9\,Mpc (see Table \ref{table:snr}).
An additional piece of information comes from comparing the samples
of Muxlow et al.~(1994) with Fenech et al.~(2008) for the remnants in 
M82.  The latter are three times more sensitive than the former, and
yet, comparing the two datasets we see that the average
SNR luminosity decreases by just a factor of $\sim1.5$, from 
$\log_{10}[\nu L_\nu^{\rm mean}]\approx35.1$ to $\approx34.9$.
A detailed comparison of the luminosity function of detected 
SNRs in all galaxies in this sample would be useful both for 
exploring the density and magnetic field distribution of the ISM
(\S\ref{section:luminosity_function}), and for a more detailed 
understanding of the selection effects at low remnant luminosity.
Finally, we note that an additional reason that the {\it lowest}
luminosity SNRs correlate with $\Sigma_g$ is that detectable remnants
must be brighter than the diffuse radio continuum of their host
galaxy, which is larger in high-$\Sigma_g$ galaxies as a result of the
Schmidt Law ($\sds\propto\Sigma_g^{1.4}$; K98;
$\sds\propto\Sigma_g^{1.7}$; Bouch\'e et al.~2007) and the FIR-radio
correlation.

From the observed luminosities in Figure \ref{fig:meanl}, we can infer
the SNR magnetic field $B_{\rm SNR}$ under the assumptions of
equations (\ref{lrad}) and (\ref{bsnr}).  To compare this magnetic
field estimate with other constraints on the {ISM} magnetic field, we
plot these results in Figure \ref{fig:mean} in terms of the ISM
magnetic field strength $B_{\rm ISM} \equiv B_{\rm SNR}/f$, i.e., the
ambient ISM field strength that would produce a SNR field strength of
$B_{\rm SNR}$ by flux freezing.  The correlation in Figure
\ref{fig:meanl} combined with equation (\ref{bsnr}) implies that
$B_{\rm SNR}$ and $B_{\rm ISM}$ are strong functions of $\Sigma_g$.  A
simple linear least-squares fit to the data points for the average
SNRs (horizontal bars) gives \beq \log_{10}\left[\frac{B_{\rm
      ISM}}{{\rm G}},\frac{B_{\rm SNR}/3}{{\rm G}}\right]\approx-3.20
+0.565\log_{10}\left[\frac{\Sigma_g}{{\rm g\,\,cm^2}}\right].
\label{fitb}
\eeq For the brightest SNRs the relation is similar, but with
normalization $-2.90$ and slope 0.52.  Note that for larger assumed
$p$, the correlation between $B_{\rm ISM}$ and $\Sigma_g$ flattens
(see the discussion after eq.~\ref{bsnr}).  Figure \ref{fig:mean}
demonstrates that for normal spirals the magnetic field strength
$B_{\rm ISM}$ inferred from the average-brightness SNR is comparable
to or larger than the equipartition field, $B_{\rm eq}$ (dotted line),
and larger than the minimum magnetic field strength $B_{\rm ph}$
(dashed line; eq.~\ref{bph}); this is true up to $\Sigma_g \sim 0.1$ g
cm$^{-2}$, which characterizes starbursts such as NGC 253 and M82.
For the densest, most luminous starbursts in our sample (Arp 299 and
the nuclei of Arp 220), $B_{\rm ISM}$ is below $B_{\rm eq}$ and --- at
least for Arp 220 --- comparable to $B_{\rm ph}$. At the other 
extreme, for the lowest density systems in our sample, NGC 300 and NGC 4736,
Figure \ref{fig:mean} tentatively implies a trend of larger 
$B_{\rm ISM}/B_{\rm eq}$ at low $\Sigma_g$.  

Because of the paucity of systems at very high $\Sigma_g$ in Figures
\ref{fig:meanl} and \ref{fig:mean}, we searched the literature for
additional high resolution radio observations of local starbursts and
ULIRGs. Unfortunately, the existing observations are not yet at
sufficiently high spatial resolution to unambiguously imply that
single SNRs have been detected.  For example, Momjian et al.~(2006)
found 10 compact radio sources in the ULIRG IRAS 17208-0014.  Most
have deconvolved sizes at half maximum of $\sim10$\,pc. At a distance
of $\approx170$\,Mpc ($z=0.043$), it is unclear if these compact
sources are indeed individual SNRs, rather than unresolved populations
of multiple remnants.  In addition, the spectral indices of these
sources are unknown, and many may be bright HII regions.  A single
source is unresolved, with a linear scale $<4.9$\,pc.  This is also
the least luminous source with
$S_\nu(1.6{\rm\,GHz})=119\pm45$\,$\mu$Jy.  Given the gas surface
density from K98 ($\approx0.4$\,g cm$^{-2}$), this individual source
would lie above the single Arp 299 remnant (see Table \ref{table:snr})
in Figure \ref{fig:meanl} by a factor of roughly two.  Similarly,
recent observations of IRAS 23365+3604 and IRAS 07251-0248 have been
reported by Romero-Ca{\~n}izales et al.~(2008).  They reveal a number
of compact radio components.  However, because these galaxies are at
distances of $\approx230$ and $\approx360$\,Mpc, respectively, the
resolution of the observations is again insufficient to associate
individual sources with single SNRs.  Comparing with the observations
of the nuclear sources in Arp 220 by Lonsdale et al.~(2006), we expect
that the sources detected by Romero-Ca{\~n}izales et al.~(2008) are
likely to be collections of $\sim10$ SNRs per source.  Higher
resolution studies can confirm or exclude this possibility.  Taking
the sources from Romero-Ca{\~n}izales et al.~(2008) at face value, and
estimating the gas surface density of these ULIRGs from the
literature, both systems would lie significantly above the solid line
($B_{\rm eq}$) in Figure \ref{fig:mean}, and they would deviate
markedly from the trend seen in Figure \ref{fig:meanl}.

\begin{figure*}
\centerline{\psfig{file=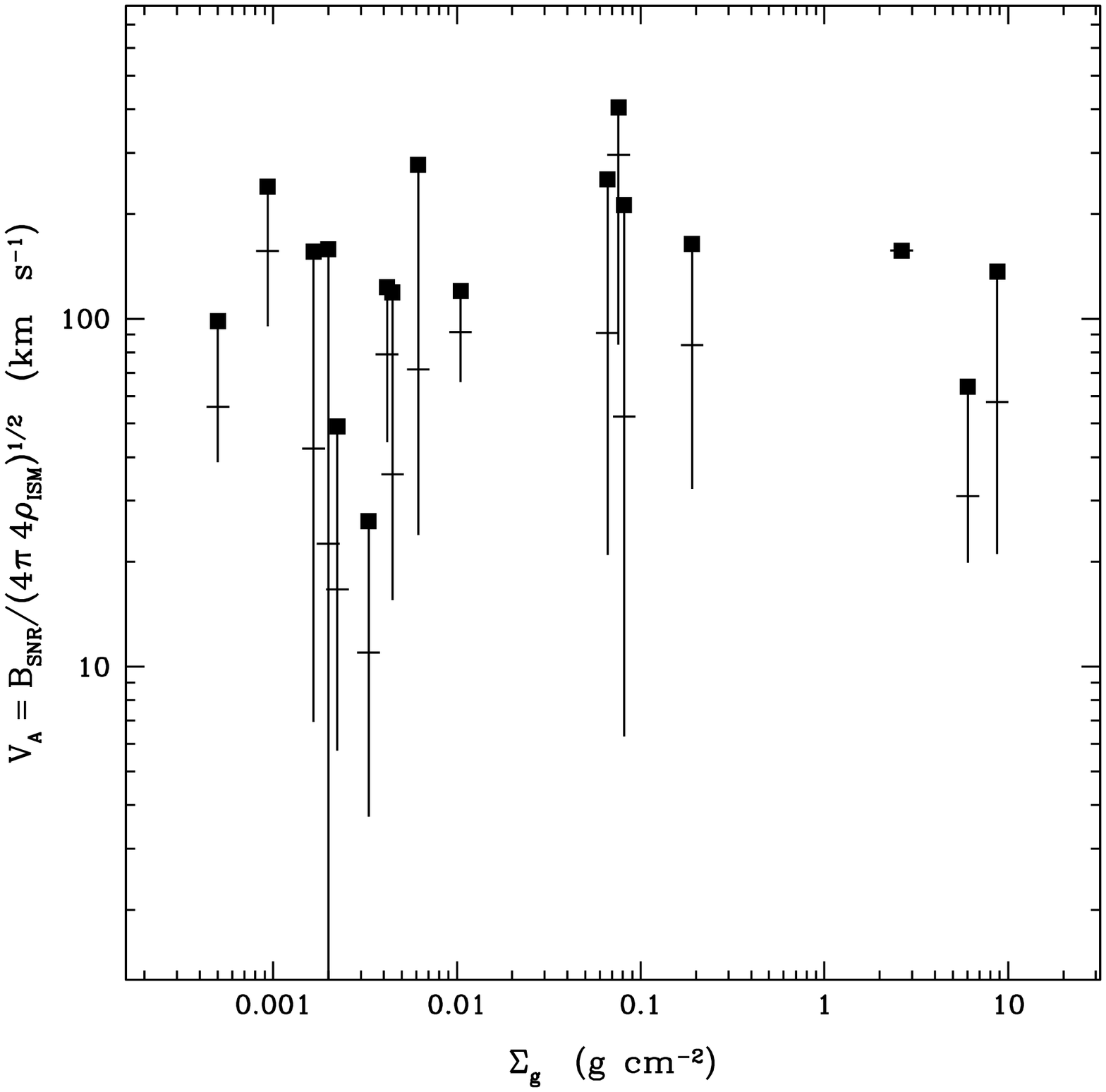,width=8.8cm}\psfig{file=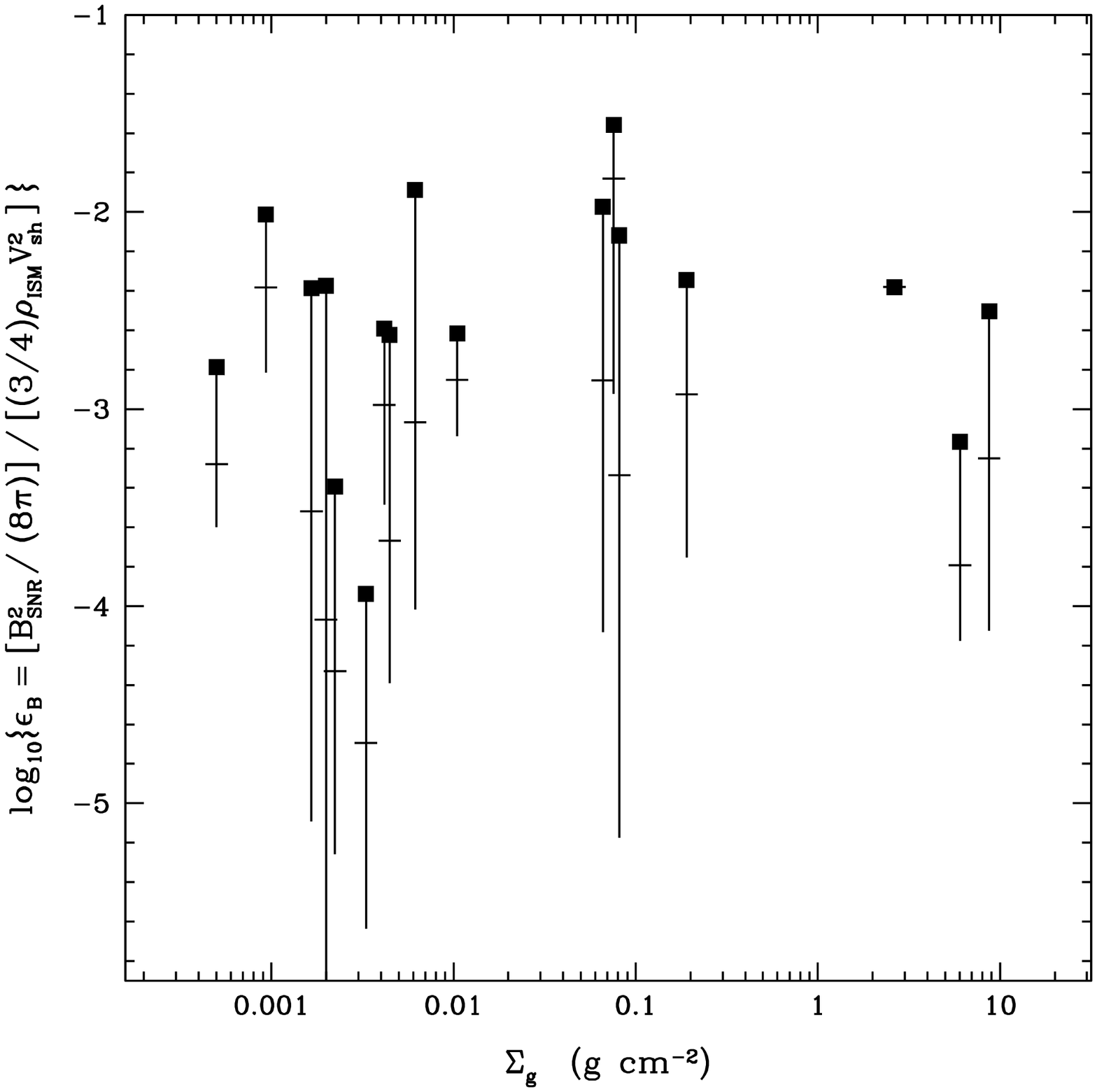,width=8.8cm}}
\caption{{\it Left Panel:} SNR Alfv\'en speed ($V_A$) as a function of the 
average ISM gas surface density $\Sigma_g$.  Here, $\rho_{\rm ISM}=\Sigma_g/(2h)$, 
where $h=100$\,pc is the assumed gas scale height for all galaxies.  Note
that the range of $\Sigma_g$ plotted here corresponds to the range of average
densities $0.5\,\,{\rm cm^{-3}}\,\lesssim n_{\rm ISM}\lesssim10^4$\,cm$^{-3}$. 
{\it Right Panel:} The ratio of the post-shock magnetic field energy density to the 
post-shock thermal pressure, $\epsilon_B=[B_{\rm SNR}^2/(8\pi)]/[(3/4)\rho_{\rm ISM}V_{\rm sh}^2]$,
as a function of $\Sigma_g$ for $h=100$\,pc.  In both panels,
as in Figure \ref{fig:mean}, $B_{\rm SNR}$ is inferred from equation (\ref{bsnr})
and the SNR radio luminosities in Table \ref{table:snr}.  Note that if supernovae
explode in regions of lower-than-average density (\S\ref{section:amplification}, 
\S\ref{section:luminosity_function}),  the inferred $V_A$ and $\epsilon_B$ both increase.
\label{fig:meann}}
\end{figure*} 

\section{Discussion}
\label{section:discussion}

Figure \ref{fig:meanl} demonstrates that SNR luminosities are a
strong, and fairly continuous, function of $\Sigma_g$, the average
(galaxy-wide) gas surface density of their host galaxy.  There may be
a change in the correlation for $\Sigma_g\gtrsim0.1$\,g cm$^{-2}$, but
more observations of SNRs in dense starbursts are required to assess
this.  In the following, we consider two plausible interpretations of
this data: (1) the luminosity of a SNR is significantly larger in
denser galaxies because the ambient magnetic field is significantly
larger (as in Fig. \ref{fig:mean}), or (2) SNRs are brighter in denser
galaxies because the magnetic field in the post-shock plasma scales as
$B^2/8\pi \propto \rho_{\rm ISM} v_{\rm sh}^2$ and is thus larger in
galaxies with a denser ISM (larger $\rho_{\rm ISM}$).

\subsection{Field Amplification}
\label{section:amplification}

Recall that $B_{\rm ph}$ (eq.~\ref{bph}, dashed lines in
Fig.~\ref{fig:mean}) is a {\it lower limit} to the true average ISM
magnetic field strength because for $B<B_{\rm ph}$ IC losses dominate
synchrotron losses and one expects a systematic deviation from the
FIR-radio correlation at high FIR luminosities, which is not observed
(T06).  In addition, the magnetic field inferred from SNRs, $B_{\rm
  ISM}$ (eq.~\ref{bsnr}, points in Fig.~\ref{fig:mean}) is an
approximate {\it upper limit} to $B$ because post-shock turbulence and
instabilities can amplify magnetic fields in SNRs (V\"olk et al.~2002;
Berezhko et al.~2006).  It is striking that the upper and lower limits
on $B_{\rm ISM}$ coincide to within a factor of few for high
$\Sigma_g$ starburst galaxies, particularly given the maximum
equipartition magnetic fields these galaxies could support ($B_{\rm
  eq}$, eq.~\ref{beq}).  This correspondence implies that shock
compression of $B_{\rm ISM}$ is sufficient to power SNR radio
emission in dense starburst galaxies.

For the specific case of Arp 220, our inferred ISM magnetic field
strength of $\sim 1-3$\,mG is also reasonably consistent with the
direct Zeeman detections of Robishaw et al.~(2008); thus three
independent observational methods, each likely probing somewhat
different phases of the ISM, all imply that a magnetic field of a few
mG pervades the nuclear region of Arp 220.  A corollary of this is
that Figure \ref{fig:mean} provides strong evidence for the
interpretation of the radio point sources in Arp 220 as normal SNRs
expanding into a highly-magnetized ISM; this alleviates the need for
{\it bona fide} radio SNe to generate the very high luminosities
observed (Smith et al.~1998; Rovilos et al.~2005; Parra et al.~2007;
see also Chevalier \& Fransson 2001).

In contrast to starbursts like Arp 220, in normal spirals independent
observations constrain the ambient ISM magnetic field to be $\sim 5-10
\, \mu$G (e.g., Beck 1982; Fitt \& Alexander 1993; Beck \& Krause
2005).  Compressing this field by a factor of $\sim 3-7$ results in a
magnetic field of sufficient strength to explain the typical radio SNR
observed in normal spirals (Fig.~\ref{fig:mean}); it appears, however,
that the brightest remnants require fields that are a factor of few
larger (and perhaps more at the lowest surface densities), either 
because the SNe explode in a region of
larger-than-average $B$ or because of modest post-shock field
amplification.  Nonetheless, as in the case of starbursts, Figure
\ref{fig:mean} implies that significant post-shock magnetic field
amplification (say $> \times10$) is not required to explain the
{\it average} radio luminosities of SNRs in spiral galaxies.

In our own Galaxy, there is some evidence for magnetic field
amplification in young shell-type SNRs (e.g., V\"olk et al.~2002;
V\"olk et al.~2005; Berezhko et al.~2006).  In these models,
individual Galactic SNRs are observed and modeled at a moment in their
evolution, in contrast to the statistical approach taken here.  For
example, V\"olk et al.~(2005) constrain $B_{\rm SNR}\sim300\,\mu$G in
Tycho's SNR by fitting the broadband X-ray and radio emission
simultaneously.  Using $\xi=1$ in equation (\ref{bsnr}), and using
only the radio data, we find that $B$ in the remnant is
$\sim9$\,$\mu$G.\footnote{This is for an adopted distance to
  the Tycho SNR of 2.4\,kpc (from the Green catalog; see Table
  \ref{table:snr}).}
We have also compared the results from V\"olk et al.~(2005) with our
calculation for the SNRs RCW 86, SN 1006, and Cass A.  We find that
$B_{\rm SNR}=3B_{\rm ISM}\approx9$, 4, and 180\,$\mu$G, whereas V\"olk
et al.\ find $\approx100$, $160$, 500\,$\mu$G, respectively.  The
difference in the inferred SNR field between V\"olk et al.\ and this
work lies partially in the assumptions about the energy in
relativistic electrons, encapsulated in our parameter $\xi$ in
equations (\ref{lrad}) and (\ref{bsnr}).  In order to not over-produce
the radio luminosity with such a large $B$, V\"olk et al.~(2006)
decrease the cosmic-ray electron-to-proton ratio, effectively using a
small value of $\xi$.  In contrast, the normalization of the FIR-radio
correlation, which we use here to normalize the emission from all
individual SNRs, requires that $\xi \simeq 1$ when averaged over a
suitably large number of remnants.  We expect that shock accelerated
electrons will have a total energy equivalent to $\xi \simeq 1$ by the
onset of the Sedov-Taylor phase, when the shocked, swept-up mass is
comparable to the ejecta mass.  If shock acceleration only becomes
efficient very late in a remnant's evolution --- or if the electrons
rapidly diffuse out of the remnant on a timescale comparable to the
Sedov-Taylor time --- then the assumption of $\xi \simeq 1$ may be
unreasonable for a typical remnant.  Although more definitive
calculations are required to assess these possibilities, we regard
both of them as unlikely and thus conclude that the majority of
observed remnants can be interpreted assuming $\xi \simeq 1$.  In this
case, although individual remnants, at particular moments in time, may
have fields larger than that due to compression alone, Figure
\ref{fig:mean} suggests that this is not a significant effect for SNRs
in starbursts, or for the average SNR in most normal spirals.

A SNR could in principle generate magnetic fields in approximate
equipartition with the post-shock thermal pressure of the gas: 
$B_{\rm SNR}^2/8\pi \sim \rho_{\rm ISM} v_{\rm sh}^2$ where $\rho_{\rm ISM}$
is the density of the ISM in which the SN explodes and $v_{\rm sh}$ is
the velocity of the SN shock at a given time (this assumption is often
used in radio SNe and $\gamma$-ray burst models; e.g., Chevalier 1982;
Meszaros \& Rees 1997); see Figure \ref{fig:bshock} for estimates of
the magnetic field strengths in SNRs if significant post-shock field
amplification indeed occurs.  In this interpretation, the higher
remnant luminosities in denser galaxies (Fig.~\ref{fig:meanl}) are due
to the post-shock scaling $B_{\rm SNR} \propto \rho_{\rm ISM}^{1/2}$,
rather than anything about the magnetic field in the ambient ISM. 

Adopting the interpretation that post-shock magnetic field
amplification is solely responsible for generating $B_{\rm SNR}$ and
the observed radio emission, Figures \ref{fig:meanl} and
\ref{fig:mean} have several interesting implications.  Taking a
constant gas scale height of $h\approx100$\,pc for all galaxies, so
that the average ISM density is $\rho_{\rm ISM} = n_{\rm ISM}m_p =
\Sigma_g/(2h)$, Figure \ref{fig:mean} implies that $B_{\rm
  SNR}\propto\rho_{\rm ISM}^{1/2}$, i.e., there is a constant
characteristic Alfv\'en speed in all observed remnants.\footnote{
Although the
assumption of constant gas scale height across all galaxies in the
sample is a reasonable first approximation, they may have
systematically smaller $h$ at higher $\Sigma_g$, which would tend to
flatten the function $V_A(n_{\rm ISM})$ (see eq.~\ref{fitva}).  }  
The left panel of Figure \ref{fig:meann} shows this explicitly. Here, we have
plotted $V_A=B_{\rm SNR}/(4\pi 4\rho_{\rm ISM})^{1/2}$ as a function
of $\Sigma_g$, assuming a constant gas scale height.  A fit to the
average SNRs gives 
\beq \log_{10}\left[\,\frac{V_A}{{\rm
      km\,\,s^{-1}}}\,\right]\approx1.87 +0.065\log_{10}\left[\frac{\Sigma_g}
{{\rm g\,\,cm^{-2}}}\right].
\label{fitva}
\eeq For the brightest remnants, the normalization is higher
($\approx2.17$) and the slope is flatter ($\approx0.021$).  
In Figure \ref{fig:meann}, $V_A$ is calculated assuming that the SNR
density is $4\rho_{\rm ISM}$, appropriate for a strong adiabatic
($\gamma = 5/3$) shock expanding into the ISM.  If, as we discuss in
\S\ref{section:luminosity_function}, supernovae tend to sample low
density regions of the ISM, the implied $V_A$ would increase, probably
by a factor of $\sim 3-10$.  Despite uncertainty in the overall
normalization for $V_A$, the left panel of Figure \ref{fig:meann}
shows that the Alfv\'en speed within remnants is roughly the same for
all observed SNRs in all galaxies; if magnetic field amplification is
at work in SNRs, this finding is an important clue to the physics.

This result can equivalently be presented as a constraint on
$\epsilon_B$, the ratio of the post-shock magnetic energy density to
the post-shock pressure (see Fig.~\ref{fig:bshock}).  The right panel
of Figure \ref{fig:meann} shows our determination of $\epsilon_B$ for
the SNRs in our sample as a function of the average ISM gas density.
For the average SNRs, we find that \beq
\log_{10}\left[\epsilon_B\right]\approx-3.04
+0.131\log_{10}\left[\frac{\Sigma_g}{{\rm g \,\, cm^{-2}}}\right].
\label{fiteps}
\eeq For the brightest remnants, the normalization is considerably
higher ($\approx-2.44$) and the slope is yet flatter ($\approx0.042$).
The value of $\epsilon_B$ is proportional to $V_A^2$, so the
near-constancy of $\epsilon_B$ with $n_{\rm ISM}$ is not surprising
given the results for $V_A$ in the left panel of Figure
\ref{fig:meann}.  As with $V_A$, we emphasize that the precise value
of $\epsilon_B$ depends sensitively on whether or not supernovae
explode in gas at the average ISM density.  If, as is more likely,
most of the volume is filled with gas 10 to 100 times less dense than
$n_{\rm ISM}$ (\S \ref{section:luminosity_function}), the typical
inferred value of $\epsilon_B$ would be $\sim 0.01-0.1$.

Because the observations do not definitively favor either SN shock
compression of ISM field or post-shock field amplification,
we are unable to unambiguously determine which of these two
possibilities is correct.  An important testable difference between
these two hypotheses lies in the dependence of the radio emission on
shock velocity or SNR diameter $D_{\rm SNR}$.  For shock compression
alone, $\nu L_\nu$ should be independent of $v_{\rm sh}$ while if
$B^2/8\pi \propto \rho_{\rm ISM} v_{\rm sh}^2$, $\nu L_\nu \propto
v_{\rm sh}^{3/2}$.  Another way to state this result is that the two
models predict different SNR surface brightness-to-diameter relations
(the $\Sigma_{\rm SNR}-D_{\rm SNR}$ relation).  For shock compression
alone, $\Sigma_{\rm SNR} \propto D_{\rm SNR}^{-2}$ while for $B^2
\propto v_{sh}^2$, $\Sigma_{\rm SNR} \propto D_{\rm SNR}^{-17/4}$ (in
both cases assuming $p = 2$; see also, e.g., Berezhko \& V\"olk 2004).
Observations of SNRs in the Milky Way find $\Sigma_{\rm SNR} \propto
D_{\rm SNR}^{-2.4}$ (Case \& Bhattacharya 1998), much closer to the
flux-freezing prediction, while observations of SNRs in M82 find
$\Sigma_{\rm SNR} \propto D^{-3.5}$ (Huang et al.~1994), closer to the
predictions of the field amplification model (see Urose{\v v}i\'c et
al.~2005).  For M82, the recent work by Fenech et al.~(2008) finds a
best fit of $\Sigma_{\rm SNR} \propto D^{-3.0}$.  Updated
observational samples, particularly those that characterize SNRs in
high-density regions, and more sophisticated theoretical models are
needed to fully understand the implications of the $\Sigma_{\rm
  SNR}-D_{\rm SNR}$ relation and its connection to the correlation
presented in Figure \ref{fig:meanl}.

\subsection{The Dynamical Role of Magnetic Fields}

Another important conclusion from Figure \ref{fig:mean} is that
$B_{\rm ISM} \ll B_{\rm eq}$ in the most luminous starbursts.  This
implies that magnetic fields are not dynamically important for
hydrostatic equilibrium on large scales in these systems, in sharp
contrast with spiral galaxies like our own.  One possible explanation
for this difference is that the lifetime of intense starbursts ($\sim
10^7-10^8$ yrs) is significantly shorter than that of the relatively
continuous star formation in normal spiral galaxies; there may thus be
less time for dynamo processes to amplify the average ISM magnetic
field.

Figure \ref{fig:mean} also suggests that the magnetic energy density
$U_B$ is comparable to the photon energy density $U_{\rm ph}$ in a
wide range of systems, with $U_B \sim {\rm few-}10 \, U_{\rm ph}$, depending
on the form of the Schmidt Law adopted (compare the two dashed lines
in Fig.~\ref{fig:mean}).  This result is not as surprising as it
might first appear because massive stars determine both the radiation
field of galaxies and the energy and momentum injected into their ISM.
The latter ``feedback'' can generate turbulence in the ISM, amplifying
the magnetic field.  Thompson et al.~(2005) and Thompson (2008) have
suggested a direct connection between turbulence in dense starbursts
and $U_{\rm ph}$ by positing that radiation pressure from the
absorption and scattering of stellar light by dust grains dominates
the overall pressure support.  The rapid diffusion of infrared photons
in starbursts causes the medium to be gravitationally unstable
(Thompson 2008) and, even in the absence of self-gravity, generates
magnetohydrodynamic instabilities that can amplify magnetic fields
directly (Turner et al.~2007).  In addition, the momentum supplied to
the ISM by stellar winds and SN explosions is comparable to that in
photons.  There are thus a number of sources capable of generating
turbulent motions that could amplify magnetic fields to the levels
inferred here.  It is also worth noting that although $B_{\rm ph} \ll
B_{\rm eq}$ for luminous starbursts in Figure \ref{fig:mean}, this does not
necessarily imply that radiation pressure is dynamically unimportant;
the dense gas in starbursts is optically thick even in the far
infrared and thus the energy density of photons in the densest gas is
larger than that in equation (\ref{bph}) by a factor of the dust
optical depth (see footnote 1), which can be $\sim 10-100$ (e.g.,
Thompson et al. 2005).

\subsection{SNR Luminosity Function}
\label{section:luminosity_function}

As a final comment, we note that a more sophisticated understanding of
the SNR luminosity function is needed to provide a more quantitative
interpretation of the data in Figures \ref{fig:meanl} \&
\ref{fig:mean}.  For example, although equation (\ref{lrad}) suggests
that $\nu L_\nu$ is independent of time, the radio luminosity of a
given SNR will peak at its Sedov time and then decline when the
remnant cools and the shock loses energy (the radio luminosity can
decline more quickly if $t_{\rm syn}$ is short or if the relativistic
electrons rapidly diffuse out of the remnant; see, e.g., van
der Laan 1962; Baring et al.~1999;
Berezhko \& V\"olk 2004).  If the average ISM
magnetic field is volume-filling, and if SNe sample the medium fairly,
taking an average density of Arp 220 of $\langle n\rangle
\sim10^4$\,cm$^{-3}$ and a SN rate of $\sim1$\,yr$^{-1}$, we estimate
that $\sim10^3$ SNRs should be visible in the radio, which is at odds
with the $\sim50$ SNRs identified (e.g., Lonsdale et al.~2006).
However, there is an important set of mitigating effects.  First,
calculations of high Mach number isothermal turbulence imply that for
the conditions of Arp 220 the average medium is filled with gas with
$n\sim10^{-2}\langle n\rangle\approx10^2$\,cm$^{-3}$ (e.g., LeMaster
\& Stone 2008).  Second, although the synchrotron cooling time $t_{\rm
  syn}$ is longer for lower ambient density (because of an assumed
smaller $B_{\rm ISM}$), and one might thus expect to see more
remnants, the SNRs at $t \sim t_{\rm syn}$ are also physically larger:
for $n=10^2$\,cm$^{-3}$ a remnant expands to $\approx4$\,pc by $\sim
t_{\rm syn}$.
For Arp 220, this is large enough that the SNRs would
overlap on the sky and be resolved out by current VLBI observations.
The radio luminosity function of SNRs is thus a convolution of the
spatial variation of the ISM conditions into which SNe explode (e.g.,
variations in $\rho_{\rm ISM}$ and $B_{\rm ISM}$), the time-dependence of an
individual remnant, and the sensitivity and resolution of the observation.
A careful study of these effects might provide important constraints
on the nature of the ISM --- the statistics of the density field and
its magnetization --- in external galaxies.  In addition,
it would more quantitatively determine the importance of 
post-shock magnetic field amplification in SNRs, and how the 
average (or maximum) SNR luminosity is related to the 
host galaxy's volume-averaged ISM magnetic field strength.

\section*{Acknowledgments}

We thank Anatoly Spitkovsky, Boaz Katz, Roger Chevalier, 
Heinrich V\"olk, 
and the anonymous referee for comments that improved  this paper. 
T.A.T.~thanks Jos\'e L.~Prieto, Mark Krumholz, Scott Gaudi, and 
Paul Martini for useful conversations, Axel Weiss for providing 
data from M82 used to compute the average column density, L.~Maddox
for detailed discussion on radio sources in M51, and 
the Aspen Center  for Physics where a portion of this work was completed.  
E.Q. is supported in part by NASA grant NNG06GI68G and the David and
Lucile Packard Foundation.  N.M.~is supported in part by a Canadian
Research Chair in Astrophysics.


\onecolumn
\newpage

\begin{table}
\begin{center}
\caption{Radio Emission from Supernova Remnants \label{table:snr}}
\begin{tabular}{lcccccccccc}
\hline \hline

\\

\multicolumn{1}{c}{System} & 
\multicolumn{1}{c}{$D$$^a$} & 
\multicolumn{1}{c}{$N_{\rm SNR}$$^b$} &
\multicolumn{1}{c}{$\Sigma_g$$^c$} & 
\multicolumn{1}{c}{$\log[\nu L_\nu^{\rm mean}]$\,\,$^d$}&
\multicolumn{1}{c}{$\log[\nu L_\nu^{\rm max}]$\,\,$^e$}&
\multicolumn{1}{c}{Freq.$^f$} &
\multicolumn{1}{c}{Beam$^g$} &
\multicolumn{1}{c}{Sensitivity$^h$} & 
\multicolumn{1}{c}{Refs.} &
\multicolumn{1}{c}{Refs.}\\

\multicolumn{1}{c}{Name} &
\multicolumn{1}{c}{(Mpc)} &
\multicolumn{1}{c}{} & 
\multicolumn{1}{c}{$({\rm g\,cm^{-2}})$} &
\multicolumn{1}{c}{(1.4 GHz)} &
\multicolumn{1}{c}{(1.4 GHz)} &
\multicolumn{1}{c}{(GHz)} &
\multicolumn{1}{c}{} &
\multicolumn{1}{c}{} &  
\multicolumn{1}{c}{SNRs} &
\multicolumn{1}{c}{$\Sigma_g$} \\

\multicolumn{1}{c}{} &
\multicolumn{1}{c}{} &
\multicolumn{1}{c}{} & 
\multicolumn{1}{c}{} &
\multicolumn{1}{c}{(ergs s$^{-1}$)} &
\multicolumn{1}{c}{(ergs s$^{-1}$)} & 
\multicolumn{1}{c}{} &
\multicolumn{1}{c}{} &
\multicolumn{1}{c}{} & 
\multicolumn{1}{c}{} &
\multicolumn{1}{c}{} \\

\\

\hline

& &   \\

  NGC 300 & 1.9$^p$   & 17     & 0.00050    & 33.4 & 33.7 & 1.45  & $4.7\arcsec\times3.6\arcsec$          & 60\,$\mu$Jy       & 1     &   2\\ 
 NGC 4736 & 4.4       & 10     & 0.00093    & 34.3 & 34.5 & 1.45  & \dots$^q$                             & \dots$^q$         & 3     &   3\\ 
      LMC & 0.053     & 21     & 0.0017$^i$ & 33.6 & 34.4 & 0.48  & 4\arcmin.3                            & few\,mJy          & 4,5   &   6\\ 
Milky Way & \dots$^n$ & 38$^o$ & 0.0020     & 33.2 & 34.5 & \dots & \dots                                 & \dots             & 7     &   8\\ 
      M33 & 0.84      & 51     & 0.0022     & 33.1 & 33.8 & 1.42  & 7\arcsec                              & 50\,$\mu$Jy       & 9     &   10\\ 
      SMC & 0.065     & 23     & 0.0033$^j$ & 32.9 & 33.5 & 2.37  & 40\arcsec                             & 0.4\,mJy          & 11    &   6\\ 
 NGC 6946 & 5.6       & 15     & 0.0042     & 34.3 & 34.6 & 1.4   & 2\arcsec                              & 20\,$\mu$Jy       & 12    &   10\\ 
 NGC 1569 & 2.2$^s$   & 23     & 0.0044     & 33.8 & 34.6 & 1.49  & $1.4\arcsec\times1.4\arcsec$          & 21\,$\mu$Jy       & 13    &   10\\ 
      M51 & 6.5       & 37     & 0.0062     & 34.4 & 35.2 & 1.4   & $1.50\arcsec\times1.21\arcsec$        & 22.5\,$\mu$Jy     & 14    &   10\\ 
      M83 & 7.3       & 4      & 0.011      & 34.7 & 34.9 & 1.4   & $3.5\arcsec\times3.5\arcsec$$^r$      & $45$\,$\mu$Jy$^r$ & 15    &   10\\ 
  NGC 253 & 3.2       & 11     & 0.067$^k$  & 35.3 & 35.9 & 5.0   & $0.6\arcsec\times0.3\arcsec$          & 40\,$\mu$Jy       & 16    &   10,17\\
 NGC 2146 & 14.5      & 3      & 0.076      & 36.1 & 36.3 & 1.6   & $0.19\arcsec\times0.15\arcsec$        & 35\,$\mu$Jy       & 18    &   10,19\\ 
      M82 & 3.6       & 37     & 0.081$^l$  & 34.9 & 35.9 & 5.0   & $35-50$\,mas                          & 17\,$\mu$Jy       & 20    &   21\\ 
 NGC 4945 & 3.8       & 8      & 0.19$^m$   & 35.6 & 36.0 & 2.3   & $16-15$\,mas                          & 75\,$\mu$Jy       & 22    &   23\\
  Arp 299 & 41        & 1      & 2.6        & 36.6 & 36.6 &       & $5.6\,{\rm mas}\times 4.5\,{\rm mas}$ & 35\,$\mu$Jy       & 24    &   10\\ 
 Arp 220E & 78        & 20     & 6.0        & 36.1 & 36.5 & 1.65  & $5.9\,{\rm mas}\times 2.7\,{\rm mas}$ & $5-6$\,$\mu$Jy    & 25    &   26\\ 
 Arp 220W & 78        & 29     & 8.7        & 36.6 & 37.1 & 1.65  & $5.9\,{\rm mas}\times 2.7\,{\rm mas}$ & $7-9$\,$\mu$Jy    & 25    &   26\\ 

\\

\hline
\hline

\end{tabular}
\end{center}

\vspace*{0.1cm}

$^a$Distance in Mpc.  \\
$^b$Number of supernova remnants.\\
$^c$Average gas surface density.\\
$^d$Mean SNR luminosity.  Each SNR luminosity is calculated assuming $F_\nu\propto\nu^{-0.5}$. \\
$^e$Maximum SNR luminosity. \\
$^f$Frequency of quoted flux density in SNR reference used to calculate SNR luminosity. \\
$^g$Measure of beamsize quoted in SNR reference. \\
$^h$Measure of sensitivity quoted in SNR reference.  Typically RMS sensitivity per beam. \\
$^i$Calculated using a total gas mass of $6\times10^8$\,M$_\odot$ (Israel 1997) 
         and $R_{25}\approx4.9$\,kpc. \\
$^j$Calculated using a total gas mass of $4.5\times10^8$\,M$_\odot$ (Israel 1997) 
         and $R_{25}\approx3.0$\,kpc. \\
$^k$Surface density scaled to K98, but with the $X_{\rm CO}$ conversion factor 
         advocated in Mauersberger et al.~(1996a).\\
$^l$Total gas mass of starburst region $2.3\times10^8$\,M$_\odot$ (Weiss et al.~2001), 
         with diameter $\approx870$\,pc, adjusted for $D=3.6$\,Mpc. \\
$^m$Total gas mass within a radius of 12\arcsec ($\approx221$\,pc at $D=3.8$\,Mpc) 
         is taken as $1.5\times10^8$\,M$_\odot$ (Mauersberger et al.~1996b). \\
$^n$All distances from Green (2006); see also Case  \& Bhattacharya (1998). \\ 
$^o$From Green catalog.  Only ``S''-type sources with unambiguous distances are 
included. For comparison, using a sub-sample of the 20 SNRs from Case \& Bhattacharya (1998) 
that overlap with the Green catalog, we have compared the highest and average SNR luminosities 
using distances from the former and the latter, respectively.
Both the highest SNR luminosity  (Cass A) and the average SNR luminosity are 
unchanged: $\log_{10}[\nu L_\nu^{\rm mean} ({\rm 1.4\,GHz; ergs \,\,s^{-1}})]\approx34.5$ and $\approx33.5$,
respectively.  However, for the full Green catalog of 38 sources (listed above), 
$\log_{10}[\nu L_\nu^{\rm mean} ({\rm 1.4\,GHz; ergs \,\,s^{-1}})]$ decreases to $33.2$,
a factor of  $\approx2.0$ lower than the smaller sample of Case \& Bhattacharya (1998).\\
$^p$Distance from Gieren et al.~(2005). \\
$^q$Values not listed in reference. \\
$^r$See Maddox et al.~(2006), their Table 1. The four sources listed are those with confirmed optical counterparts. \\
$^s$Distance from Israel (1988).\\

References: 
(1) Pannuti et al.~(2000); see also Payne et al.~(2004); 
(2) Read et al.~(1997);
(3) Duric \& Dittmar (1988); 
(4) Mathewson et al.~(1983); 
(5) Mathewson et al.~(1984); 
(6) Israel (1997);
(7) fluxes from Green, D.~(2006) (see {\tt http://www.mrao.cam.ac.uk/surveys/snrs/}); see also Case \& Bhattacharya (1998); 
(8) Boulares \& Cox (1990);
(9) Gordon et al.~(1999); 
(10) K98; 
(11) Filipovi{\'c} et al.~(2005);
(12) Hyman et al.~(2000); 
(13) Chomiuk \& Wilcots (2009); see also Greve et al.~(2002);
(14) Maddox et al.~(2007), including all sources steeper than $\alpha=0.4$,
excluding sources 53 (the nucleus) and 104 (background; Ho \& Ulvestad 2001); 
(15) Maddox et al.~(2006);
(16) Ulvestad \& Antonucci (1997), see also Ulvestad \& Antonucci (1991); 
(17) Mauersberger et al.~(1996a);
(18) Tarchi et al.~(2000); 
(19) Greve et al.~(2006); 
(20) Fenech et al.~(2008), but see also Kronberg et al.~(1985), Huang et al.~(1994),
 Muxlow et al.~(1994); Wills et al.~(1997), and Allen \& Kronberg (1998); 
(21) Weiss et al.~(2001);
(22) Lenc \& Tingay (2008)
(23) Mauersberger et al.~(1996b); 
(24) Neff et al.~(2004);
(25) Lonsdale et al.~(2006), and see
also Rovilos et al.~(2005), Parra et al.~(2007); 
(26) Downes \& Solomon (1998)

\label{lastpage}

\end{table}

\end{document}